\documentclass[preprint2]{aastex}
\usepackage{aastexug}

\def\plotfiddle#1#2#3#4#5#6#7{\centering \leavevmode
\vbox to#2{\rule{0pt}{#2}}
\includegraphics{#1}}

\def\eqalign#1{\null\,\vcenter{\openup\jot 
	\ialign{\strut\hfil$\displaystyle{##}$&$
	\displaystyle{{}##}$\hfil \crcr#1\crcr}}\,}

\begin{document}

\title{Galactic Bulge Pixel Lensing Events}

\author{Cheongho Han}
\affil{Department of Physics,
Chungbuk National University, Chongju 361-763, Korea}
\email{cheongho@@astroph.chungbuk.ac.kr}

\begin{abstract}
Gould \& DePoy proposed a pixel lensing survey towards the Galactic bulge 
using a small aperture ($\sim 65$ mm) camera with a large pixel size ($\sim 
10''$) detector and deliberately degraded optics achieving $30''$ point 
spread function (PSF).  In this paper, we estimate the event rate of this 
pixel lensing survey expected under various detection criteria, which are 
characterized by the threshold signal-to-noise ratio, $(S/N)_{\rm th}$, 
and event duration, $t_{\rm dur,th}$, and investigate the characteristics 
of the detectable events.  From this investigation, we find that the event 
rate varies significantly in the range $\sim 6\ {\rm yr}^{-1}$ -- $120\ 
{\rm yr}^{-1}$ depending strongly on the imposed detection criteria, 
implying that to maximize event detections it will be essential to identify 
events by diligently inspecting light variations and to promptly conduct 
followup observations for the identified events.  Compared to events 
detectable from classic lensing surveys, the events detectable from the 
pixel lensing survey will generally involve brighter source stars and have
higher amplifications. For the pixel lensing events detectable under the 
criteria of $(S/N)_{\rm th}=10$ and $t_{\rm dur,th}=6$ hr, we find that
the baseline brightness of source stars will be in average $\sim 2$ mag 
brighter than those of classic lensing events and $\sim 90\%$ will have 
amplifications $A\geq 20$ and $\sim 40\%$ will be extreme microlensing 
events (EMEs) with $A\geq 200$.  Therefore, followup observations of the 
pixel lensing events will provide high quality data, which enable one to 
precisely determine the lensing parameters and obtain extra-information 
about the lenses and source stars.  Especially, high amplifications events 
with $A\geq 20$ will be important targets for high-efficiency planet 
detections and one can uniquely determine the mass, distance, and transverse 
speed of individual lenses for EMEs.  
\end{abstract}

\section{Introduction}
The previous and the current microlensing surveys towards the Galactic 
bulge and the Magellanic Clouds \citep{alcock1993, aubourg1993, udalski1993, 
alard1997, abe1997} have been and are being carried out by using the 
conventional photometric method based on the extraction of the individual 
source stars' point spread function (PSF).  With this method, the 
lensing-induced light variations can be measured only for events with 
resolved source stars, and thus events occurred to unresolved faint 
sources cannot be detected.  Pixel lensing, on the other hand, detects 
and measures the variation of a source star flux by subtracting successive 
images of the same field from a normalized reference image \citep{tomaney1996, 
alard1998, alard1999}, enabling one to measure the light variations even 
for unresolved source stars.  The pixel lensing technique was originally 
proposed to search for lensing events towards the unresolved star field 
of M31 \citep{crotts1992, baillon1993}. Recently, pixel lensing searches 
have been conducted towards the resolved Galactic bulge field 
\citep{alcock1999a, alcock1999b, alcock2000}, but they are intended to 
make use of the additional sources lying below the detection threshold, 
and so to increase the sensitivity of the existing experiments.

\citet{gould1998} proposed a radically different type of pixel lensing 
survey towards the Galactic bulge using a small aperture ($\sim 65$ mm) 
camera with a large pixel-size ($\sim 10''$) detector and deliberately 
degraded optics achieving $30''$ PSF.  With these instruments, the 
observations are analogous to the normal pixel lensing observations of 
the bulge of M31, but are carried out under conditions where the detected 
events can be followed up in detail.  Since the proposed PSF set by the 
optics will not vary in time with atmospheric conditions, the data 
reduction process will be very simple.  Compared to classic lensing 
surveys based on PSF photometry, the proposed pixel lensing survey has 
an advantage of the broad coverage of most or all of the ($\sim 50\ 
{\rm deg}^2$) southern bulge field of interest with the large field of 
view ($\sim 36\ {\rm deg}^2$ with a 2K$\times$2K CCD camera).  In addition, 
the instrumentation required to carry out the survey will be inexpensive.  
\citet{gould1998} described the possibility of the Galactic bulge pixel 
lensing survey in general, but detailed estimation of the event rate and 
investigation about the characteristics of the detectable events have not 
been done.

In this paper, we estimate the event rate of the proposed Galactic pixel 
lensing survey expected under various detection criteria 
and investigate the characteristics of the detectable events.

\section{Event Rate Estimation}

\subsection{Event Production}
To estimate the event rate, we perform simulations of Galactic 
bulge pixel lensing events by producing a large number of artificial light 
curves.  The light curve of a pixel lensing event is represented by
\begin{equation}
F= F_0(A-1), 
\end{equation}
\begin{equation}
A={u^2+2\over u\sqrt{u^2+4}};\ \ \
u = \left[ \beta^2+ \left({t \over t_{\rm E}}\right)^2\right]^{1/2},
\end{equation}
where $F_0$ is the baseline flux of the source star, $A$ is the amplification,
$u$ is the lens-source separation (normalized by the angular Einstein ring
radius $\theta_{\rm E}$), $t_{\rm E}$ is the Einstein ring radius crossing
time (Einstein timescale), and $\beta$ is the impact parameter (also
normalized by $\theta_{\rm E}$).  The Einstein timescale is related to the
physical parameters of the lens system by
\begin{equation}
t_{\rm E}={r_{\rm E}\over v};\ \ \ 
r_{\rm E}=D_{\rm ol}\theta_{\rm E}
         =\left[ {4Gm\over c^2} 
          {D_{\rm ol}(D_{\rm os}-D_{\rm ol})\over D_{\rm os}}
          \right]^{1/2},
\end{equation}
where $r_{\rm E}$ is the Einstein ring radius, $m$ is the lens mass, $v$ 
is the lens-source transverse speed, and $D_{\rm ol}$ and $D_{\rm os}$ 
represent the distances to the lens and source from the observer, 
respectively.  Therefore, to produce a pixel lensing event light curve, 
it is required to define the lensing parameters of $F_0$, $\beta$, $m$, 
$D_{\rm os}$, $D_{\rm ol}$, and $v$.

We produce light curves by selecting the lensing parameters so that the 
event rate, $\Gamma$, becomes 
\begin{equation}
\eqalign{
\Gamma \propto
    & \int dD_{\rm os}\ \rho(D_{\rm os}) \int dD_{\rm ol}\ \rho(D_{\rm ol}) 
      \int\hskip-2pt\int dv_y dv_z\ vf(v_y,v_z)\cr
    & \int dm\ \Phi(m)\ r_{\rm E} \int dL\ \Phi(L) \int d\beta\ f(\beta),\cr
}
\end{equation}
where $\rho(D_{\rm ol})$ and $\rho(D_{\rm os})$ are the density distributions 
of the lens and source locations, $(v_y,v_z)$ are the two components of the 
transverse velocity, $f(v_y,v_z)$ is their distribution, $\Phi(m)$ is the 
lens mass function, $\Phi(L)$ is the absolute luminosity function of source 
stars, and $f(\beta)$ is the distribution of impact parameters. In eq.\ (4), 
the factors $r_{\rm E}$ and $v=({v_y^2+v_z^2})^{1/2}$ are included because 
events with larger cross-sections (i.e.\ $r_{\rm E}$) and higher transverse 
speeds have higher chances of being lensed.

For the source and lens density distributions, we adopt a barred bulge 
model of \citet{dwek1995} and a double-exponential disk model of 
\citet{bahcall1986}.  The Dwek et al.\ bulge model is represented by
\begin{equation}
\rho = \rho_0\ 
\exp \left[ -0.5 r_s^2 \right]\ M_\odot/{\rm pc}^3,
\end{equation}
where $r^4=[(x'/x_0)^2+(y'/y_0)^2]^2+(z'/z_0)^4$, $(x',y',z')$ are the three 
axes of the triaxial bulge ($x'$ is the longest and $z'$ is the shortest), 
and $(x_0,y_0,z_0)=(1.58, 0.62, 0.43)\ {\rm kpc}$ are the scale lengths. The 
normalization constant $\rho_0$ is set so that the total bulge mass becomes 
$\sim 1.8\times 10^{10}\ M_\odot$ \citep{han1995}.  The Bahcall disk model 
is represented by
\begin{equation}
\rho = 0.06 \exp\left[ -\left( {R-R_0\over h_R}+{z\over h_z}\right)
\right]\ M_\odot/{\rm pc}^{3},
\end{equation}
where $h_R=3.5\ {\rm kpc}$ and $h_z=325\ {\rm pc}$ are the radial and vertical 
scale heights of the disk and $R_0=8\ {\rm kpc}$ is the galactocentric 
distance of the sun.

For the transverse velocity distribution, we adopt the model of 
\citet{han1995}.  In this model, the velocity distributions for both disk 
and bulge components have a Gaussian form, i.e.\ 
\begin{equation}
f(v_i) \propto \exp \left[ -{(v_i-\bar{v}_i)^2\over 
               2\sigma_i^2}\right];\ \ \  i \in y,z,
\end{equation}
where the means and the standard deviations of the individual velocity
components are
\begin{equation}
(\bar{v}_y,\sigma_y)=
\cases{
(220.0,30.0)\ {\rm km\ s}^{-1} & (disk)\cr
(0.0,82.5) \ {\rm km\ s}^{-1}  & (bulge),\cr
}
\end{equation}
\begin{equation}
(\bar{v}_z,\sigma_z)=
\cases{
(0.0,20.0)\ {\rm km\ s}^{-1} & (disk)\cr
(0.0,66.3)\ {\rm km\ s}^{-1} & (bulge).\cr
}
\end{equation}

\begin{figure}[t]
\plotfiddle{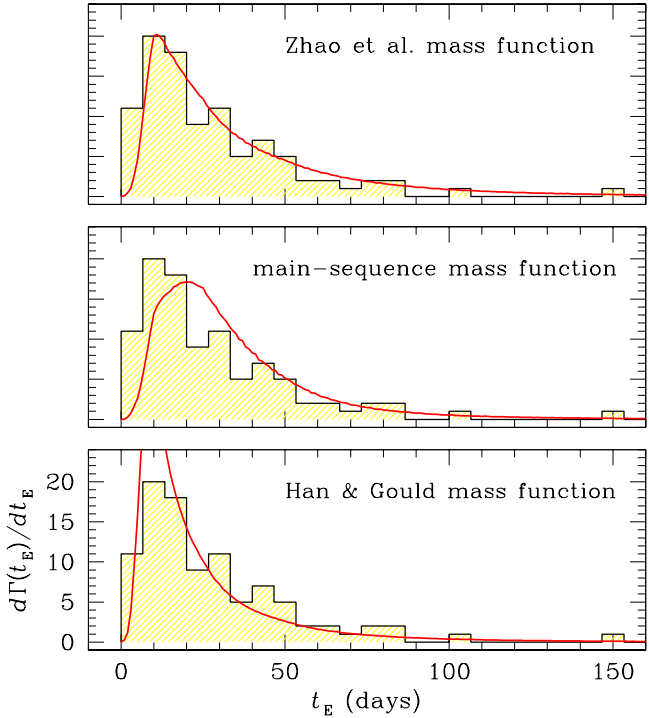}{0.0cm}{0}{48}{48}{-150}{-330}
\vskip9.2cm
\caption{Comparison of the Einstein timescale distributions resulting from
three different model mass functions (solid curve in each panel) with the 
distribution based on the latest data of the 97 candidate bulge events 
detected by the MACHO group (histogram).  The theoretical distributions are 
corrected by the detection efficiency and normalized so that the total 
number of events  matches the observed one. }
\end{figure}

Unlike the models of the matter density and velocity distributions that are
constrained by observations, our knowledge about the lens mass function 
is very poor.  Therefore, we start with several proposed models of the 
lens mass function and compute the resulting $t_{\rm E}$ distributions,
$d\Gamma(t_{\rm E})/dt_{\rm E}$, combined with the adopted models of the 
matter density and velocity distributions.  Then the mass function model 
producing $t_{\rm E}$ distribution that best matches the observed one is 
used for our simulation.  We test three mass function models.  These 
models are a main-sequence mass function of \citet{kroupa1993} (with 
three powers of $\alpha=-4.5$ for $m>1.0\ M_\odot$, $\alpha=-2.5$ for 
$0.5\ M_\odot < m \leq  1.0\ M_\odot$, and $\alpha=-1.2$ for $0.08\ M_\odot 
< m \leq  0.5\ M_\odot$), a single power-law distribution of \citet{han1996} 
(with $\alpha =-2.3$ and a mass cutoff of $m_{\rm cut}=0.04\ M_\odot$), 
and another power-law distribution of \citet{zhao1995} (with $\alpha =-2.0$ 
and $m_{\rm cut}=0.08\ M_\odot$).  In Figure 1, we present the distributions 
of Einstein timescales resulting from the model mass functions and they 
are compared with that of the latest $t_{\rm E}$ data of the 97 events 
detected by the MACHO group \citep{alcock1999b, alcock2000}.  One finds 
that the $t_{\rm E}$ distribution resulting from the Zhao et al.\ (1995) 
model is consistent with the observed one, while those resulting from the 
\citet{han1996} and Kroupa et al.\ (1993) models produce short timescale 
events that are more than and less than the observed one, respectively.  
Therefore, we adopt the Zhao et al.\ (1995) model for our simulation.

For the source brightness distribution, we adopt the $I$-band absolute 
luminosity function determined by \citet{kroupa1995}.  We assume that 
photometry of the pixel lensing survey is carried out in $I$ band because 
more light can be obtained due to its large band width and small amount of 
extinction compared to other optical bands.  Once the location of a source 
star and its absolute brightness, $M_I$, are selected, the apparent source 
star brightness is obtained by calculating the corresponding distance 
modulus, $\mu$, and extinction, $A_I$, i.e.\ $I=M_I + \mu(D_{\rm os})+
A_I(D_{\rm os})$.  Due to the patchy and irregular characteristics of the 
Galactic interstellar absorbing matter distribution, it is difficult to 
model its detailed distribution.  Hence, we approximate that the 
interstellar matter is distributed similar to the Galactic disk matter 
distribution.  With this approximation, the source flux is decreased from 
its de-reddened value by a factor $\exp \left[ -\kappa\Sigma (D_{\rm os})
\right]$, where $\Sigma (D_{\rm os})$ represents the column density of 
the absorbing matter located between the observer and source star and 
$\kappa$ is a normalization constant.  \citet{stanek1996} determined that 
the extinction in $V$ band towards the Baade's Window (BW) varies 
$A_V\sim 1.26$ -- $2.79$ with the most frequent value of $\langle A_V\rangle 
\sim 1.7$.  With the relation $A_V/E(V-I) \sim 2.49$ also provided by him, 
this corresponds to $A_I\sim 0.76$ -- $1.67$ with $\langle A_I\rangle 
\sim 1.02$.  Therefore, we set the value $\kappa$ so that the average 
amount of extinction towards the BW becomes $A_I=1.02$ at a distance of 
$D_{\rm os}=8\ {\rm kpc}$.  We discuss the effect of variable extinction 
with position on the sky in the following subsection.

Under the definition of a lensing event as `a close lens-source encounter 
within the Einstein ring of a lens', the impact parameters are randomly 
chosen, i.e.\ $f(\beta)\equiv {\rm constant}$, in the range $0\leq \beta
\leq 1$.

\subsection{Efficiency Determination}
Once events are produced, we then determine the efficiency of pixel lensing
event detection by imposing the following detection criteria.

First, we require detectable events should have high signal-to-noise ratios 
during their peak amplifications.  Since the uncertainty of the light 
variation measurement of a pixel lensing event is almost totally dominated 
by the background flux, the signal-to-noise ratio is computed by 
\begin{equation}
S/N \sim F_0(A-1)\left( {t_{\rm exp}\over F_0 A+\langle B\rangle}\right)^{1/2},
\end{equation}
where $t_{\rm exp}=2$ min is the mean exposure time and $\langle B\rangle$ 
is the average background flux.  Determination of $\langle B\rangle$ is 
described below.

Second, since event detection is also restricted by the duration of events, 
$t_{\rm dur}$, we additionally require detectable events should last long 
enough for solid confirmation.  We define the duration as `the length of 
the time period during which the signal-to-noise ratio exceeds a threshold 
value of $(S/N)_{\rm th}$' and it is computed by
\begin{equation}
t_{\rm dur} = 2t_{\rm E}\left(\beta_{\rm th}^2-\beta^2\right)^{1/2},
\end{equation}
where $\beta_{\rm th}$ represents the maximum allowed impact parameter for 
an event to satisfy the threshold signal-to-noise ratio \citep{han1999}.  
To compute $\beta_{\rm th}$, we first compute the threshold amplification 
$A_{\rm th}$ corresponding to $(S/N)_{\rm th}$ by numerically solving 
eq.\ (10).  With the obtained value of $A_{\rm th}$, then, the threshold 
impact parameter is obtained by
\begin{equation}
\beta_{\rm th} = \left[ 2(1-A_{\rm th}^{-2})^{-1/2}-2\right]^{1/2}.
\end{equation}
With these detection criteria, the detection efficiency is determined by 
obtaining the ratio of the number of events satisfying the above detection 
criteria out of the total number of tested events.

\begin{figure}[t]
\plotfiddle{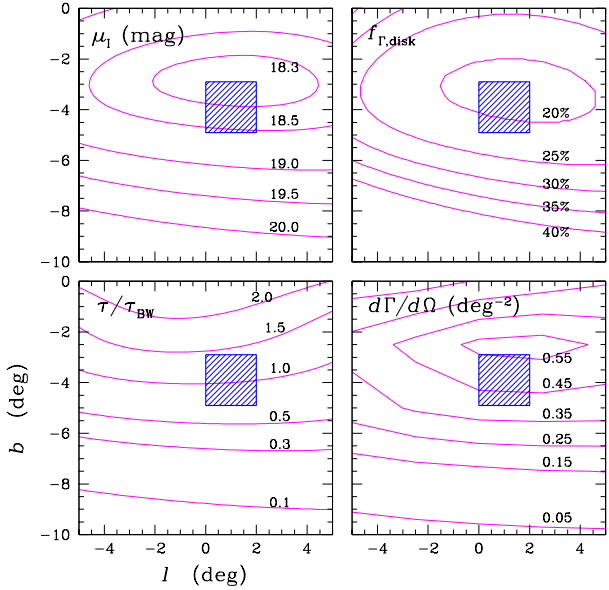}{0.0cm}{0}{42}{42}{-130}{-282}
\vskip7.5cm
\caption{
Maps of the surface brightness, $\mu_I$, optical depth, $\tau$,
disk event fraction, $f_{\Gamma,{\rm disk}}$, and pixel lensing event
rate, $d\Gamma/d\Omega$ of the Galactic bulge field. The shaded 
square represents the region around the Baade's Window, which is centered 
at $(l,b) \sim (-1.0^\circ,-3.9^\circ)$. The event rate map is for the 
pixel lensing events detectable under the criteria of $(S/N)_{\rm th}=10$ 
and $t_{\rm dur,th}=6$ hr.}
\end{figure}

\begin{figure}[t]
\plotfiddle{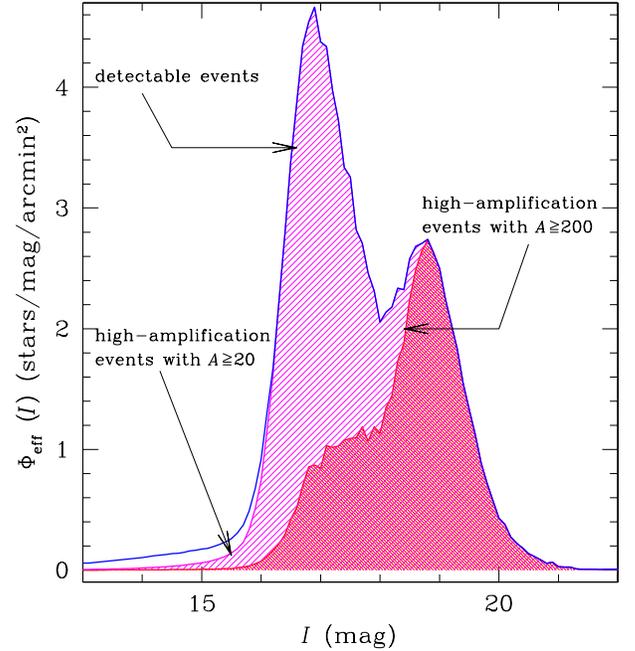}{0.0cm}{0}{48}{48}{-150}{-315}
\vskip8.4cm
\caption{ The distribution of baseline source brightnesses for pixel events 
(effective luminosity function). The presented effective luminosity 
function is for the source stars of the pixel lensing events detectable 
towards the BW under the detection criteria of $(S/N)_{\rm th}=10$ and 
$t_{\rm dur,th}=6$ hr. Also presented are the effective luminosity functions 
for events with high amplifications: the light shaded region for events with 
$A\geq 20$ and the dark shade region for events with $A\geq 200$.  }
\end{figure}

The proposed pixel lensing survey will be performed on a large area of 
the sky.  Then, the event rate per unit angular area, $d\Gamma/d\Omega$, 
will be different for different regions of the sky because the stellar 
distribution is not uniform due to the variation both in geometric 
distance from the Galactic center and extinction. In addition, the spatial 
variation of the optical depth, $\tau$, and the disk/bulge event ratio 
with the position on the sky also makes the event rate depend on the 
observed field. To consider this effect, we construct maps of the surface 
brightness, $\mu_I(l,b)$, optical depth, $\tau(l,b)$, and the disk event 
fraction, $f_{\Gamma,{\rm disk}}(l,b)$. The surface brightness map is 
constructed by using the matter density distribution and extinction models 
described in \S\ 2.1.  The map is normalized so that the de-reddened 
surface brightness of the BW field contributed by bulge stars becomes 
$\sim 17.6\ {\rm mag}\ {\rm arcsec}^2$ \citep{terndrup1998} and the amount 
of the extinction towards the same field becomes $A_I=1.02$ mag.  With the 
additional flux from disk stars, which is $\sim 25\%$ of the flux from 
bulge stars, and the sky background flux, which contributes $\sim 20\%$ 
of the total stellar background flux, then the average background flux 
within the PSF (with an angular area of $\Omega_{\rm PSF}\sim 0.75\ 
{\rm arcmin}^{-2}$) of the BW field becomes $\sim 800\ {\rm e}^{-1}\ 
{\rm s}^{-1}$.  The optical depth represents the average probability for 
a single source to be gravitationally amplified by a factor $A\geq 1.34$ 
at a given moment and it is computed by
\begin{equation}
\eqalign{
\tau = & \int_0^\infty dD_{\rm os}\ \rho(D_{\rm os}) 
         \int_0^{D_{\rm os}}\ dD_{\rm ol}\ 
         \rho (D_{\rm ol})\ \pi r_{\rm E}^2 \cr
       & \left[ \int_0^\infty dD_{\rm os}\ \rho(D_{\rm os})\right]^{-1}.\cr
}
\end{equation}
The disk event fraction is approximated by $f_{\Gamma,{\rm disk}}\sim\eta 
f_{\tau,{\rm disk}}$, where $f_{\tau,{\rm disk}}$ represents the ratio of 
disk event contribution to the optical depth and $\eta\sim 0.7$ is included 
because disk lensing events tend to have longer $t_{\rm E}$ \citep{han1995}
and thus $f_{\Gamma,{\rm disk}}$ is smaller than $f_{\tau,{\rm disk}}$. In 
Figure 2, we present the maps of $\mu_I(l,b)$, $\tau(l,b)$, and $f_{\Gamma,
{\rm disk}}(l,b)$.

In Figure 3, we present the baseline source star brightness distribution 
for pixel lensing events (effective source luminosity function, 
$\Phi_{\rm eff}$).  The presented $\Phi_{\rm eff}$ is for the source stars 
of the events detectable towards the BW under the criteria of $(S/N)_{\rm th}
=10$ and $t_{\rm dur,th}=6$ hr.  It is normalized so that the surface flux 
contributed by all source stars matches the background stellar flux.  In the 
figure, we also present the effective luminosity functions for events with 
high amplifications: light shaded region for events with $A\geq 20$ and dark 
shaded region for events with $A\geq 200$ (extreme microlensing events, EMEs).

\begin{figure}[t]
\plotfiddle{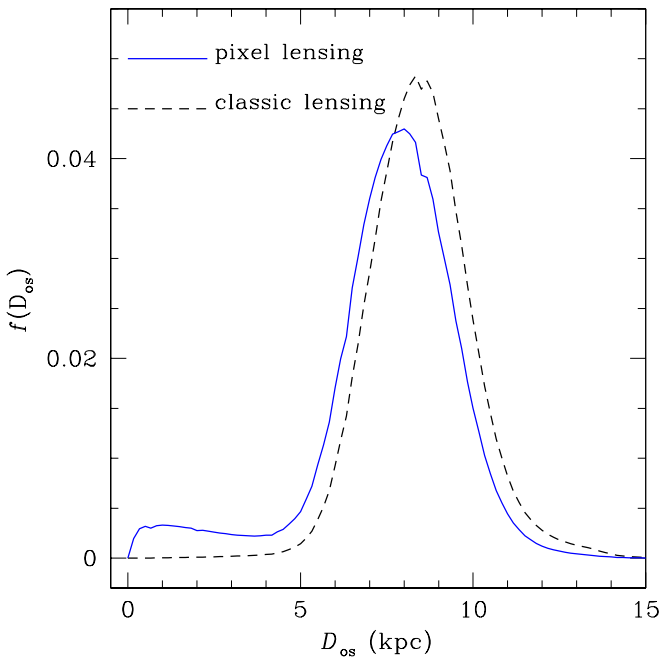}{0.0cm}{0}{48}{48}{-150}{-315}
\vskip8.4cm
\caption{
Comparison of the distributions of the source locations for pixel and classic
lensing events.  The pixel lensing distribution is for the events detectable 
towards the BW under the criteria of $(S/N)_{\rm th}=10$ and $t_{\rm dur,th}
=6$ hr.  The distributions are arbitrarily normalized.
}
\end{figure}

\subsection{Total Event Rate}
With the obtained effective luminosity function, the total pixel lensing 
event rate is then calculated by
\begin{equation}
\eqalign{
 & \Gamma_{\rm tot}=\int_{\Omega_{\rm tot}} {d\Gamma\over d\Omega}\ d\Omega;\cr
 & {d\Gamma\over d\Omega}={\pi\over 2}t_{\rm obs} N_{\rm eff} \tau_{\rm p}
	       \left<{1\over t_{\rm E}}\right>,\cr
}
\end{equation}
where 
$N_{\rm eff}=\int \Phi_{\rm eff}(I)\ dI$ is the total number of effective 
source stars of a field, $\langle 1/t_{\rm E}\rangle$ is the mean value of 
the inverse Einstein timescales of the events detected towards the field, and 
$\tau_{\rm p}$ is the pixel lensing optical depth.  Compared to the source 
stars of events detectable from classic lensing surveys (classic lensing 
events), those of pixel lensing events will be systematically brighter, 
and thus closer.  This can be seen in Figure 4, where we plot the 
distributions of the source locations for classic, $f_{\rm c} (D_{\rm os})$, 
and pixel lensing events, $f_{\rm p}(D_{\rm os})$.  As a result, the pixel 
lensing optical depth will be smaller than the value determined by the 
classic lensing survey, $\tau_{\rm c}$. We, therefore, apply $\tau_{\rm p}$ 
decreased from $\tau_{\rm c}$ by a factor
\begin{equation}
\eqalign{
{\tau_{\rm p} \over \tau_{\rm c}} = 
   &
    {\int\ [d\tau(D_{\rm os})/dD_{\rm os}]\ f_{\rm p}(D_{\rm os})\ 
     dD_{\rm os}\over
     \int f_{\rm p}(D_{\rm os})\ dD_{\rm os}} \cr
   & \left[ 
    {\int\ [d\tau(D_{\rm os})/dD_{\rm os}]\ f_{\rm c}(D_{\rm os})\ 
     dD_{\rm os}\over
     \int f_{\rm c}(D_{\rm os})\ dD_{\rm os}}
     \right]^{-1}. \cr
}
\end{equation}
We set the mean value of the classic lensing optical depth towards the BW 
to be $\tau_{\rm c} =2.4\times 10^{-6}$ by adopting the determination of 
\citet{alcock2000}.  In the lower right panel of Fig.\ 2, we present the 
distribution $d\Gamma/ d\Omega$ for the events detectable under the criteria 
of $(S/N)_{\rm th}=10$ and $t_{\rm dur,th}=6$ hr.  We assume that the total 
observation time per year is $t_{\rm obs}=150$ days and the total angular 
area of the monitored field is $\Omega_{\rm tot}=50\ {\rm deg}^2$.

\begin{deluxetable}{ccccc}
\tabletypesize{\small}
\tablecaption{The Galactic Bulge Pixel-lensing Event Rates}
\tablewidth{0pt}
\tablehead{
\multicolumn{2}{c}{detection criteria} &
\multicolumn{3}{c}{event rate (${\rm events}/{\rm yr}^{-1}$)} \\
\colhead{$(S/N)_{\rm th}$} &
\colhead{$t_{\rm dur,th}$} &
\colhead{\ \ \ $\Gamma_{\rm tot}$\ \ \ } &
\colhead{\ \ $\Gamma_{A\geq 20}$\ \ } &
\colhead{\ \ $\Gamma_{\rm EME}$\ \ } }
\startdata
7    &  $2^{\rm hr}$  &  119.9 & 113.4 (94.6\%) & 63.8 (53.2\%) \\
--   &  $6^{\rm hr}$  &  53.1  &  47.5 (89.4\%) & 19.3 (36.3\%)  \\
--   &  $12^{\rm hr}$ &  29.8  &  25.0 (83.9\%) & 8.1 (27.2\%)  \\
\smallskip
--   &  $24^{\rm hr}$ &  12.5  &  9.1 (72.8\%)  & 2.4 (19.2\%)  \\

10   &  $2^{\rm hr}$  &  70.1  &  66.8 (95.3\%) & 38.8 (55.3\%) \\
--   &  $6^{\rm hr}$  &  28.9  &  26.0 (90.9\%) & 11.4 (39.4\%)  \\
--   &  $12^{\rm hr}$ &  13.9  &  11.5 (82.7\%) & 4.4 (31.7\%)  \\
\smallskip
--   &  $24^{\rm hr}$ &  5.8   &  4.0 (69.0\%)  & 1.1 (19.0\%) \\
\enddata
\vskip-0.6cm
\tablecomments{
The event rates expected from the proposed Galactic pixel lensing experiments 
expected to be detected under various detection criteria, which are 
characterized by the threshold signal-to-noise ratio, $(S/N)_{\rm th}$, and 
event duration, $t_{\rm dur,th}$. The values in the columns of $\Gamma_{A
\geq 20}$ and $\Gamma_{\rm EME}$ represent the rates of high amplifications 
events with $A\geq 20$ and $A\geq 200$ and their fractions, respectively.
}
\end{deluxetable}

\section{Results}
In Table 1, we list the determined Galactic pixel lensing event rates 
expected under various criteria, which are characterized by $(S/N)_{\rm th}$ 
and $t_{\rm dur,th}$.  The important results from the determined detection 
rates of pixel lensing events and their characteristics are summarized as 
follows.

First, the total event rate varies significantly in the range 
$\sim 6\ {\rm yr}^{-1}$ -- $120\ {\rm yr}^{-1}$ 
depending strongly on the imposed detection criteria.  As expected, the 
event rate increases as $(S/N)_{\rm th}$ and $t_{\rm dur,th}$ decreases.  
The strong dependency of the detection rate on $t_{\rm dur,th}$ implies that 
majority of events will have short durations.  Therefore, to maximize the 
event rate it will be essential to identify pixel lensing events by 
diligently inspecting light variations with signals as low as possible and 
to conduct followup observations for the identified events as promptly 
as possible.  If events are identified for all signals with $S/N\gtrsim 7$ 
and followup observations are carried out within an hour (corresponding 
to $t_{\rm dur,th}=2$ hr) after the identification, we find that the 
event rate can reach up to $\Gamma_{\rm tot}\sim 120\ {\rm yr}^{-1}$.

Second, the majority of pixel lensing events will have high amplifications.  
For example, we estimate that $90\%$ of the pixel lensing events detectable 
under the criteria of $(S/N)_{\rm th}=10$ and $t_{\rm dur,th}=6$ hr will 
have amplifications $A\geq 20$ and $40\%$ will be EMEs.  These fractions 
increase as the imposed  $t_{\rm dur,th}$ becomes shorter.  As will be 
discussed in \S\ 4, events with $A\gtrsim 20$ are excellent followup 
monitoring targets for high efficiency extra-solar planet detections 
\citep{griest1998, han2001}.  In addition, followup observation of EMEs 
enables one to uniquely determine the masses, distances, and transverse 
speeds of the individual lenses \citep{gould1997, sumi2000}.

\begin{figure}[t]
\plotfiddle{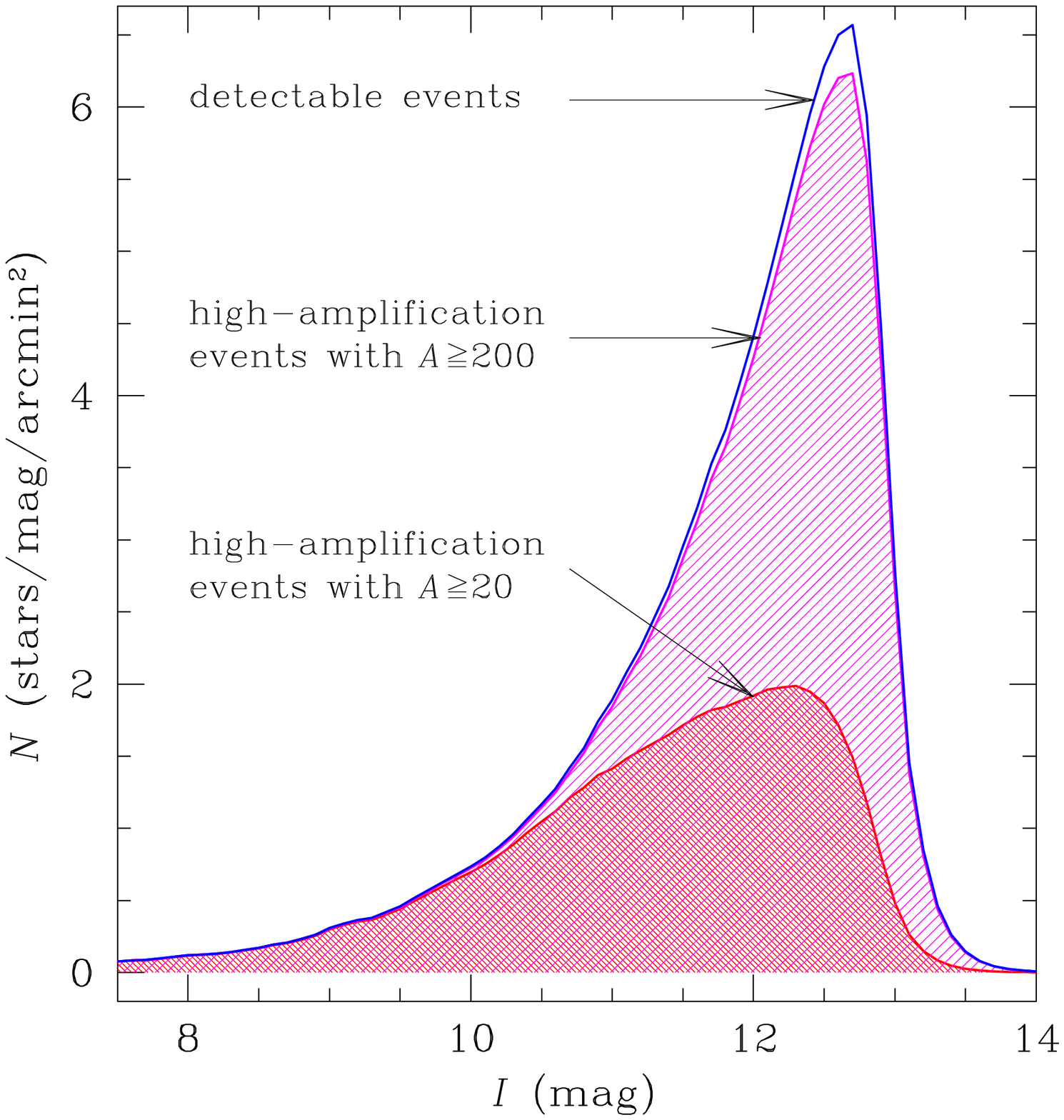}{0.0cm}{0}{48}{48}{-150}{-315}
\vskip8.4cm
\caption{
The peak brightness distribution of Galactic pixel lensing events.  The
detection criteria are same as those adopted for the computation of the 
effective luminosity function presented in Fig.\ 3.}
\end{figure}

Third, compared to classic lensing events, the events detectable by the 
proposed pixel lensing experiment will generally involve brighter source 
stars, as predicted by \citet{gould1998}.  Under the detection criteria of 
$(S/N)_{\rm th}=10$ and $t_{\rm dur,th}=6$ hr, for example, we find that 
the baseline brightness distribution of the pixel lensing event source stars
peaks at $I\sim 17$ (see Fig.\ 3), which is $\sim 2$ mag brighter than 
that of the classic lensing events [cf.\ Fig.\ 7 of \citet{alcock2000}].  
The source star brightness becomes even brighter as one applies stricter 
detection criteria, i.e.\ higher $(S/N)_{\rm th}$ and longer $t_{\rm dur,th}$.
Since pixel lensing events tend to have high amplifications, the source 
during its peak amplification will become even brighter.  In Figure 5, we 
present the distribution of peak brightnesses of source stars for events 
expected to be detected under the conditions of $(S/N)_{\rm th}=10$ and 
$t_{\rm dur,th}=6$ hr.  One finds that the peak brightnesses of nearly 
all events are brighter than $I\sim 14$.

\section{Discussion}
Due to the difference in the characteristics of pixel lensing events from 
classic ones, the proposed pixel lensing experiment has both advantages and 
disadvantages.  The greatest disadvantage is that the pixel lensing survey 
should be carried out in parallel with followup observations to obtain 
meaningful results.  However, several groups \citep{albrow1998, rhie1999} 
are already conducting such followup observations, and thus the pixel 
lensing observations can be made in parallel with these existing 
observations.  Once this problem is resolved, the proposed pixel lensing 
survey will be able to provide the following important scientific rewards.

First, one can determine the lensing parameters of events with improved 
precision and accuracy.  Precise determination of the lensing parameters is 
achieved because high precision photometry is possible for pixel lensing 
events because they generally involve brighter source stars and have higher 
amplifications.  The accuracy, on the other hand, is improved because the 
effect of blended light from unresolved nearby faint stars is minimized 
with the increased signal from the lensed source.  With the reduced 
uncertainties in the determined lensing parameters, one can better constrain 
the nature of Galactic lenses.

Second, one can obtain various useful extra information both about the 
lens and source star by efficiently detecting lensing light curve anomalies.
A lensing light curve deviates from the standard form due to various 
reasons, e.g.\ the extended source effect \citep{schneider1992, witt1994}, 
parallax effect \citep{refsdal1966, gould1994}, binary lens effect 
\citep{schneider1986, witt1990}, planetary lens effect \citep{mao1991, 
gould1992, bolatto1994}, and binary source effect \citep{griest1992, 
han1997}.  However, the amount of these deviations is usually very small, 
and thus detecting them requires high-precision observations.  Since 
better photometry is possible for pixel lensing events, the efficiency 
of detecting light curve anomalies will be increased.

Third, a significant fraction of high amplification pixel lensing events 
will be ideal monitoring targets of high-efficiency planet detections.  
\citet{griest1998} pointed out that by intensively monitoring high 
amplification events, planets located in the lensing zone ($\sim 0.6
\theta_{\rm E}$ -- $1.6\theta_{\rm E}$) can be detected with a nearly 
100\% efficiency.  Since majority of Galactic pixel lensing events will 
have high amplifications, they will be ideal targets for intensive 
followup monitoring for high-efficiency planet detections. \citet{covone2000} 
pointed out that monitoring the pixel lensing events detectable towards 
the M31 \citep{crotts1996, ansari1997, crotts2000} will also provide a 
channel for high efficiency planets detections via microlensing.  We note, 
however, that since followup observations cannot be done for M31 pixel 
lensing events, although one might detect planet-induced anomalies in the 
light curves, it will be very difficult to determine the planet parameters 
(i.e.\ the mass ratio and separation) just based on the peak part of the 
light curve.  By contrast, since one can perform high-precision followup 
observations for Galactic bulge pixel lensing events, one can accurately 
determine the planet parameters.

Another importance of the pixel lensing survey is that one can uniquely 
determine the physical parameters of individual lenses for EMEs, which will 
comprise a significant fraction of all pixel lensing events.  
\citet{gould1997} showed that followup observations of an EME enable one 
to measure both the lens proper motion and parallax that together would 
yield individual mass, distance, and transverse-speed of the lensing object.  
The proper motion is determined by analyzing the light curve anomalies 
induced by the extended source effect which occur when the lens approaches 
very close to the source star.  The parallax is determined by observing 
the difference in the peak parts of the light curves as seen from two 
Earth observers separated by $\sim R_\oplus$.

\section{Conclusion}

We estimate the event rate of the pixel lensing survey towards the 
Galactic bulge proposed by \citet{gould1998} expected under various
detection criteria and investigate the characteristics of the detectable 
events.  From this investigation, we find that the event rate varies 
significantly in the range $\sim 6\ {\rm yr}^{-1}$ -- $120\ {\rm yr}^{-1}$ 
depending strongly on the imposed detection criteria.  This implies that 
to maximize event detections it will be essential to identify events by 
diligently inspecting light variations and to promptly conduct followup 
observations for the identified events.  Compared to classic lensing 
events, the events detectable by the proposed pixel lensing survey will 
generally involve brighter source stars and have higher amplifications.
For the pixel lensing events detectable under the criteria of
$(S/N)_{\rm th}=10$ and $t_{\rm dur,th}=6$ hr, for example, 
we find that the baseline source brightness will be $\sim 2$ mag brighter 
than those of classic lensing events and most of them ($\sim 90\%$) will 
have high amplifications with $A\geq 20$ and a significant fraction of 
them ($\sim 40\%$) will be EMEs.  The high amplification events will be 
especially important because they are excellent targets for effcient 
extra-solar planet detections.  In addition, one can uniquely determine 
the mass, distance, and transverse speed of individual lensing objects 
for EMEs.

\acknowledgements
We would like to thank A.\ Gould for carefully reading the manuscript and 
making useful comments.  We also would like to express our gratitude to 
the anonymous referee for making useful comments that contributed to 
significantly improve the paper.  This work was supported by a grant 
(2000-015-DP0449) from the Korea Research Foundation (KRF).


\clearpage

\end{document}